# Broadband dielectric permittivity tensor of muscovite for next-generation all van der Waals photonic components

M. H. Hayrapetyan[1, +], M. L. Sargsyan[1, +], D. A. Karakhanyan[1, +], A. P. Khachatryan[1], M. A. Levonyan[1], D. Litvinov[2, 3], M. Koperski[2, 3], A. V. Margaryan[1], M. Šiškins[2, 4], K. S. Novoselov[2, 3] & D. A. Ghazaryan[1, *]

[1]Laboratory of Advanced Functional Materials, Yerevan State University, Yerevan, 0025, Republic of Armenia

[2]Institute for Functional Intelligent Materials, National University of Singapore, Singapore, 117575, Republic of Singapore

[3]Department of Materials Science and Engineering, National University of Singapore, Singapore, 117575, Republic of Singapore

[4]School of Physics and Astronomy, University of Southampton, Highfield, Southampton, SO17 1BJ, United Kingdom

[+]These authors contributed equally to this work

[*]The correspondence should be addressed to: davitghazaryan@ysu.am

**We report a comprehensive determination of the broadband dielectric permittivity tensor of van der Waals (vdW) muscovite also referred to as mica, establishing it as a low-index low-loss platform for ultrathin nanophotonics. Resolving its anisotropic vibrational response and extracting accurate tensor components across broadband ultraviolet (UV) to near-infrared (NIR) spectral region, we show that vdW muscovite exhibits consistently low refractive indices negligible extinction and weak in-plane anisotropy allowing its effective treatment as a uniaxial dielectric in thin-film limits. Leveraging these properties, we design muscovite based vdW heterostructures pairing it $MoS_2$, engineering few-layer distributed Bragg reflectors (DBR) and dichroic beam splitters (DBS) with high efficiency robust optical performance in a broad NIR spectral region achieved with sub-micron thicknesses. Our findings spotlight the high significance of low-index extinctionless vdW crystals, positioning muscovite as a highly perspective atomically flat building block for next-generation, broadband, all-vdW nanophotonic components**.

Introduction

The reported studies of physical properties of muscovite family 3D crystals date back to 1901 persisting over 20$^{th}$ century and covering a variety of aspects starting from alluring dielectric properties ending up with Mohs hardness factor examinations[1–4]. The former scope covers either a demonstration of biaxial anisotropic properties at a fixed wavelengths or an extraction of low refractive indices accompanied by infinitesimal extinctions at exceedingly limited visible (Vis) spectral region windows and epoch-dictated technological accuracy[5–7], though catalogued for many of representative muscovite family crystals. After the isolation of graphene[8,9] and an uncoverage of vdW interaction based layered character of muscovite family crystals, the research in this direction escalated further into a hunt for sophisticated physical effects in either bulk vdW micro-crystals or their 2D counterparts. Recent advances include an isolation of monolayer muscovite by a micro-mechanical cleavage or intercalation-assisted liquid-phase exfoliation, a resolution of atomically flat surfaces and thickness-dependent optical contrast arising from interference effects, enabling the practical identification of its few-layer samples[10–13]. VdW muscovite was also found to exhibit opposite signs of in-plane and out-of-plane dielectric permittivity components over selected windows in mid-IR spectral region, giving rise to both types of the natural hyperbolicities[14]. The former findings complement its use as an atomically flat substrate and gate dielectric in 2D devices[15–17] proposing a systematic integration into vdW heterostructures as functional building block layers and enablement of bubble-free vdW stacks, including refined moiré structures[18]. However, despite this progress and broad recognition of muscovite vdW crystals as underexplored material platform[19–22], its dielectric functionality remains largely limited to support roles with enormous potential in UV-*to*-NIR nanophotonic applications to be yet uncovered[23,24].

In this manuscript, we methodically approach the anisotropic vibrational and optical properties of biaxial vdW muscovite presenting it as a highly prospective material in addition (or for a replacement) to a sole robust wide-bandgap dielectric crystalline vdW candidate: hexagonal boron nitride (*h*BN), to potentially be used in a diverse range of next-generation photonic architectures, including dielectric multilayers, such as reflectors, filters, optical cavities, waveguide claddings, photonic crystal structures, optical spacers and so on. Our work comprises a symmetry-based resolution of polarization-sensitive Raman modes of vdW muscovite together with its highly accurate dielectric permittivity tensor components obtained in a wide UV-*to*-NIR spectral region. We display that biaxial vdW muscovite possesses broadband low refractive indices supplemented by low in-plane anisotropy and almost indiscernible extinction across the whole studied spectral region and can be considered as a uniaxial crystal for substantially thin optical layer thicknesses. Furthermore, employing the extracted optical constants, we design and engineer all-vdW few-layer stacks operating as NIR broadband DBR and DBS components pairing muscovite with $MoS_2$, achieving controlled interference and sturdy performance for oblique incidence angles. Our findings expand an extensively limited set of low-index and loss robust vdW dielectrics setting up muscovite as a practical building block for scalable and ultrathin all-vdW nanophotonic systems.

**Results**

**Polarization-resolved vibrational characterization of van der Waals muscovite**

The crystal structure of vdW muscovite $2M_1$ belongs to monoclinic crystal system organized into $C_2/c$ space group and defined by $(K^+, Na^+)Al_2(AlSi_3O_{10})(OH)_2$ chemical formulae (see Figure 1(a)), where $Na^+$ ions partially replace $K^+$ ions as demonstrated in previous studies with the same crystal source[18]. Here, weak interlayer vdW forces are associated with presence of $K^+/Na^+$ cleavable ionic cites suggesting AlSiO monolayers to lie within *xy* plane. The low crystalline symmetry of vdW muscovite results in emergence of in-plane anisotropy of vibrational modes[25] that can be examined by polarization-dependent Raman spectroscopy (see Figure 1(b)-(d)). The Raman intensity map acquired as function of incident light's polarization shows 13 distinct Raman active modes at 100-3800 cm⁻¹ spectral region from bulk muscovite micro-crystals in a backscattering configuration (see Methods). Note the robust optical phonon mode at 3630 cm⁻¹ (450 meV) shifts corresponding to OH group vibrations[26], and serving as vivid marker for a swift identification of in-plane crystallographic axis orthogonal directions in studied samples (see Figure 1(c)). The angular dependencies of the observed modes can be further analysed within group theory framework[27] with an intensity of each mode described by $J \propto |e_s^T \alpha(j) e_i|^2$, where $\alpha(j)$ is Raman tensor corresponding to $j^{th}$ vibrational mode, and $e_S$ with $e_i$ are polarization unit vectors of the scattered and incident lights. Using the symmetry element notations of Raman tensor (see Supplementary Note 1), and the polarization unit vectors of $e_i(\vartheta) = (\cos(\vartheta), \sin(\vartheta), 0)$, $e_S^{\parallel}(\vartheta) = e_i(\vartheta)$ for incident and scattered lights in the co-polarization regime, one may obtain the symmetric $A_g$ modes as a function of incident light's polarization angle expressing their intensities as $J^{\parallel}(A_g) \propto a^2[(\sin^2(\vartheta)+(b/a)\cos(\phi_{ba})\cos^2(\vartheta))^2+((b/a)\sin(\phi_{ba})\cos^2(\vartheta))^2]$, where $d$ symmetry element and Euler's angle between incident light's and crystal's z-axis are approximated to be equal to 0 owing to an infinitesimal contribution, $\phi_{ba}$ is the phase contrast between $a$ and $b$ symmetry elements, while getting $J^{\parallel}(B_g) = 0$ for the antisymmetric $B_g$ modes. The angular dependencies of $A_g$ modes, governed by the explicit ratio of $a/b$ symmetry elements result in anisotropic four-lobe patterns, where maxima align preferentially along crystallographic axis orthogonal directions as shown in Figure 1(d).

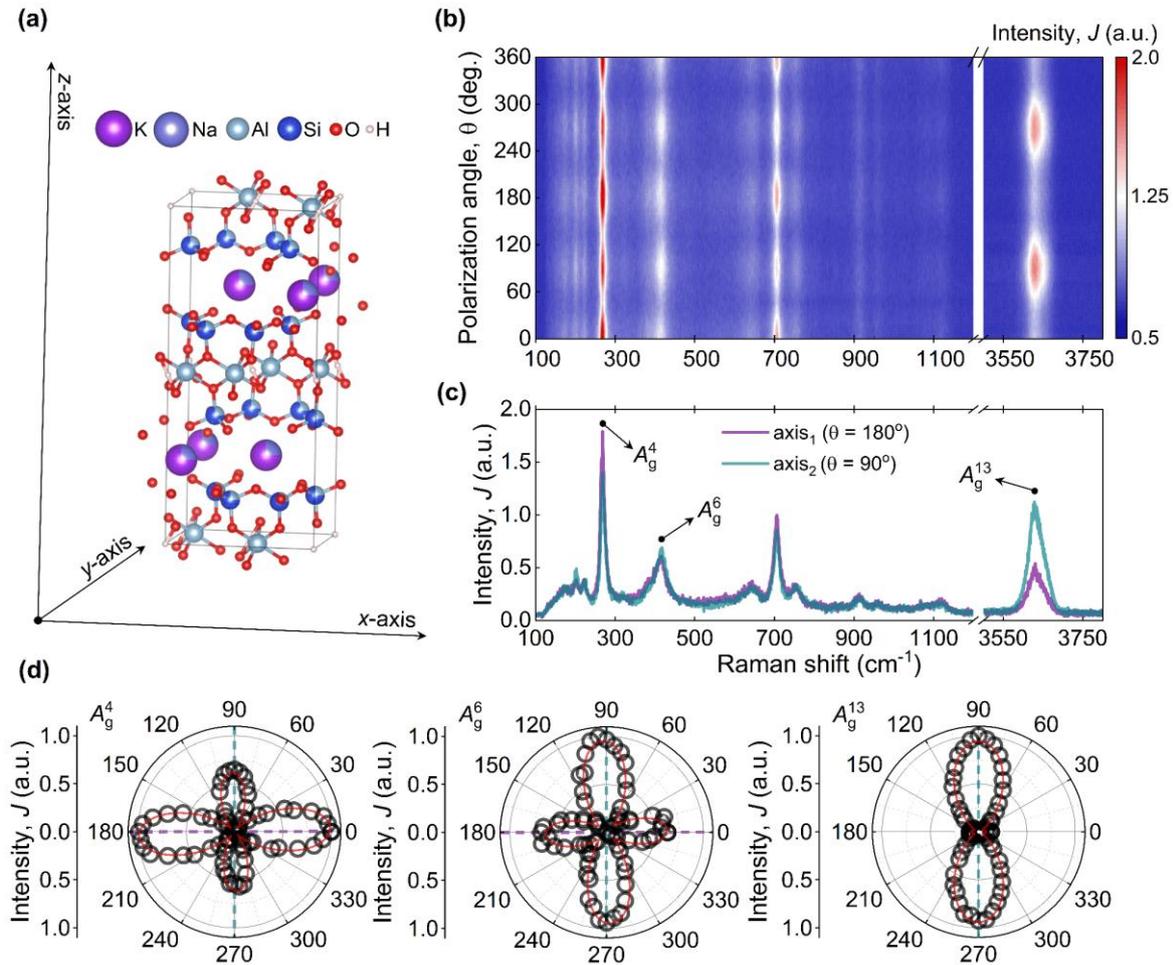

**Figure 1 | Polarization-resolved Raman spectroscopy of biaxial vdW muscovite. a,** Crystal structure of monoclinic muscovite $M_1$ generated by Vesta (formulae: $(K^+, Na^+)Al_2(AlSi_3O_{10})(OH)_2$). The crystallographic data was adopted from COD database[28] and further adapted introducing ($K^+$, $Na^+$) mixed ionic composition instead of the pristine $K^+$ ions. **b,** Polarization-resolved Raman intensity contour plot acquired in the co-polarized regime (see cross-polarized data in Figure S1) within 100-3800 cm$^{-1}$ spectral region from a bulk muscovite sample. **c,** Raman spectra acquired across in-plane crystallographic axes. **d,** Polar diagrams of $A_g$ modes illustrating the representative mode symmetries presented as function of polarization angle, $\vartheta$. Dashed lines denote the crystallographic axes of muscovite and solid curves correspond to the fits.

**Broadband dielectric permittivity tensor of low index van der Waals muscovite**

After the probing of in-plane crystallographic orthogonal axis directions, we performed spectroscopic micro-ellipsometry studies upon micro-mechanically cleaved vdW muscovite samples measuring the ellipsometric parameters $\Psi$ and $\Delta$ along them (see Methods). The latter exhibit subtle axis-dependent variations characteristic for materials with low in-plane anisotropy, turning discernible in ellipsometric parameters for thicknesses of an order of hundred nanometres (see Figure S2). To determine the complex dielectric permittivity tensor components more accurately, we employed Mueller-matrix (MM) micro-ellipsometry (see Figures S3-S4) retrieving the latter along crystallographic axes as presented in Figure 2(a). Our results display low real and imaginary dielectric permittivity tensor component parts, and thus, low refractive indices (see Figure 2(b)) together with negligible absorption in the whole Vis spectral region extended into deep UV (down to 250 nm) and NIR (up to 1800 nm) for all crystallographic directions, highlighting that vdW muscovite can be treated as an orthorhombic biaxial crystal. In particular, the extinction coefficients remain quite small even at deep UV wavelengths (*e.g.*, $k_x$ = 1.7 × 10$^{-3}$ at 250 nm) confirming the weak absorption and extra-wide-bandgap dielectric properties of muscovite[2,5]. Moreover, we continuously reproduce these ellipsometric parameters for few-layer samples without introducing any thickness-dependent variation into the optical model. This indicates that the dielectric response, and thus, the optical bandgap of vdW muscovite remain thickness-independent down to a trilayer limit (see Figure S5).

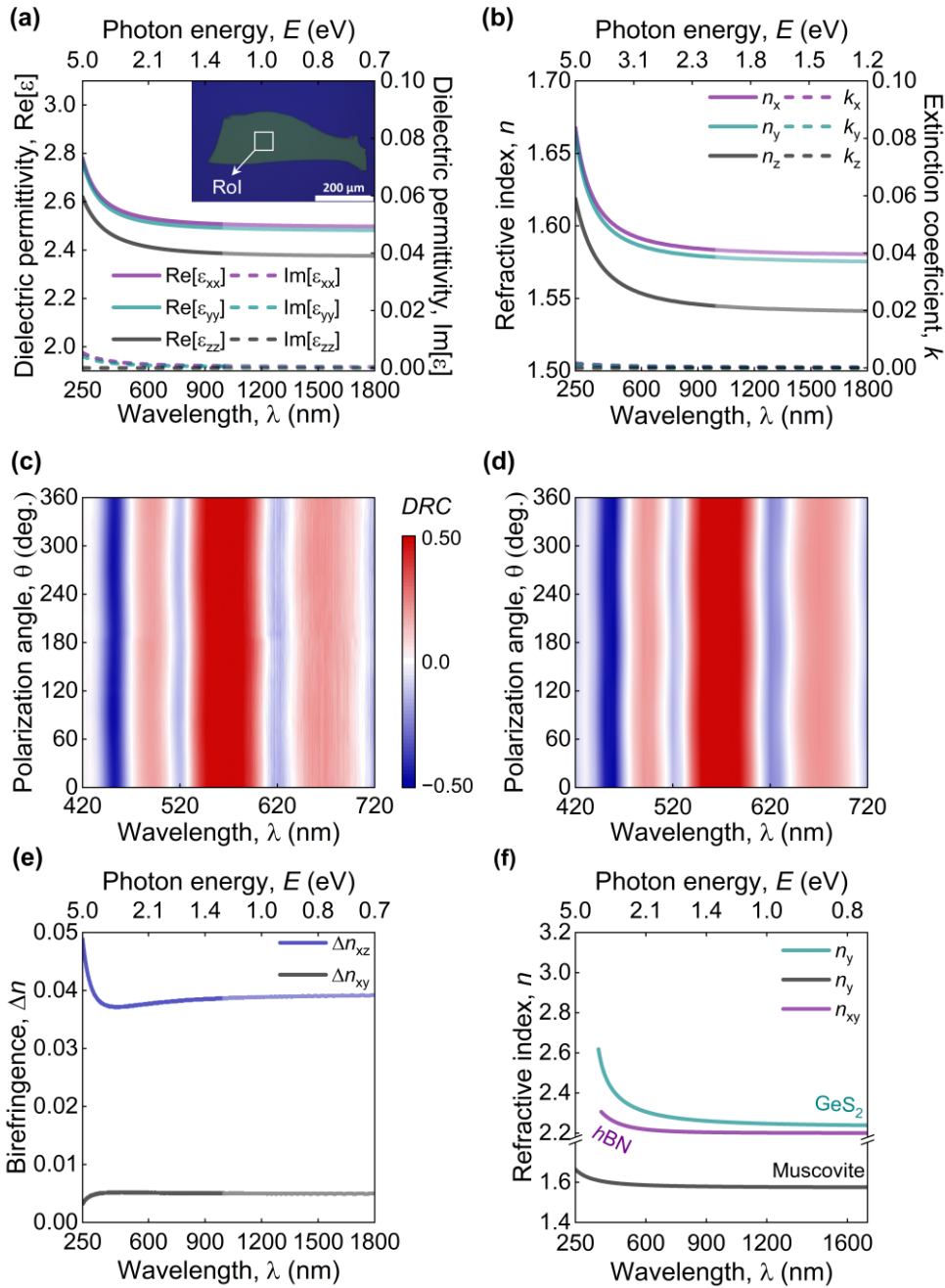

**Figure 2 | Probing the complex dielectric permittivity tensor of biaxial vdW muscovite in UV-*to*-NIR spectral region. a,** Real and imaginary (Re[$\varepsilon$] and Im[$\varepsilon$]) parts of dielectric permittivity tensor components acquired across its crystallographic axes (measured in 250-1000 nm spectral region and extrapolated up to 1800 nm as transparent curves). Inset displays the optical micrograph of a representative 815 nm thick sample used in ellipsometric (and micro-reflectance (c)) studies. RoI marks the region of data acquisition. **b,** Accurate optical constants ($n_i$, $k_i$) extracted across crystallographic axes displaying extinctionless low-refractive index properties of muscovite. The tabulated constants and the optical model of vdW muscovite are shown in Table S1-S2. **c,** Experimental polarization-resolved differential reflectance contrast (*DRC*) contour plot as function of wavelength from a representative vdW muscovite sample cleaved onto Si/SiO$_2$ substrate shown in (a). **d,** Theoretical polarization-resolved *DRC* contour plot calculated as function of wavelength using extracted optical constants shown in (b). **e,** In-plane and out-of-plane birefringences of muscovite presented as a function of wavelength. **f,** In-plane refractive index component of muscovite as a function of wavelength. Note the obtained optical dispersion positioned far beneath moderate index biaxial (or uniaxial) vdW crystals in their respective transparency windows ($k_i \leq 0.01$ for *h*BN and GeS$_2$).

To further validate our findings, we performed polarization-resolved Vis micro-reflectance spectroscopy (see Methods) for vdW muscovite samples comparing the measured dielectric response with theoretical evaluations based on the extracted optical model (see Methods), which resulted in deviations not exceeding 3 %, Figure 2(c-d) (and Figure S6). The latter confirms high accuracy of obtained dielectric permittivity components. Figure 2(e) shows the birefringent properties of vdW muscovite that exhibit small values for in-plane components, for example, $\Delta n_\parallel = 0.0051$ at 532 nm wavelengths, whereas with more pronounced out-of-plane effect $\Delta n_\perp = 0.0374$ - a tendency consistent with earlier works reporting various muscovite family crystal dielectric properties obtained in exceedingly limited spectral windows[6,7]. The former infers that sufficiently thin samples ($\leq 250$ nm) of muscovite can be treated as uniaxial with effectively isotropic in-plane permittivity components, while preserving distinct out-of-plane contrast. Notably, this low in-plane birefringence, though resolved well by micro-ellipsometry, remains challenging to detect with polarization-resolved Vis micro-reflectance spectroscopy. Furthermore, the in-plane components of the refractive indices of vdW muscovite exhibit weak dispersions remaining quite low across the studied broadband spectral region, positioning it well below the moderate-index vdW crystals[29–31], such as *h*BN and GeS$_2$ shown within their extinctionless windows (see Figure 2 (f)). For example, at 532 nm wavelengths, the latter reaches $n_y = 1.595$ underscoring muscovite's role as highly prospective broadband low-index low-loss dielectric, which can be widely employed in nanophotonic applications as a transparent dielectric media, an encapsulating layer or an optical spacer, succeeding *h*BN[32–34].

**Engineering DRBs and DBSs with van der Waals muscovite used as a low index layer**

The outstanding combination of broadband low refractive indices accompanied by negligible absorption, atomically flat interfaces and a prime compatibility for high-quality heterostructure fabrication makes vdW muscovite a unique candidate for an optical design of next-generation nanophotonic components requiring lossless operation and precise control of light-matter interaction. To leverage these advantages, we incorporate it into artificial heterostructures[35] pairing it with high refractive index MoS$_2$ counterparts for an engineering of next-generation all-vdW ultrathin broadband DBR and DBS operating at NIR spectral region, Figure 3. The schematic illustration of our few layer vdW heterostructure positioned on top of a standard fused silica substrate is displayed in Figure 3(a), where leveraging the unprecedented refractive index contrasts between muscovite and MoS$_2$ crystalline layers in the lossless NIR spectral region, we engineer a tailored optical response. Here, we first tackle traditional all-vdW DBR design with a central wavelength of 1310 nm, where the thickness of each composing layer follows the quarter-wave condition ($\lambda_c/4n$), ensuring constructive interference of the reflected lights. Our five-layer optical design including explicit refractive indices and thicknesses of constituent layers is shown in Figure 3(b). The computed reflectance (and transmittance) spectra present broadband high-reflectance (and transmittance) band spanning over 1040–1800 nm spectral region, Figure 3(d). Next, to demonstrate a proof-of-principle experiment, the corresponding vdW heterostructures composed of two muscovite and three MoS$_2$ layers were assembled on top of fused silica substrate using the deterministic dry-transfer technique (see Methods). Figure 3(d) shows the measured spectra acquired from the fabricated stack (see the optical micrograph in the inset) along with the fitted simulations. Here, 680 nm thick prototype archives the targeted performance demonstrating average reflectance exceeding 93 % above 1020 nm (the individual experimental layer thicknesses are provided in Table S3).

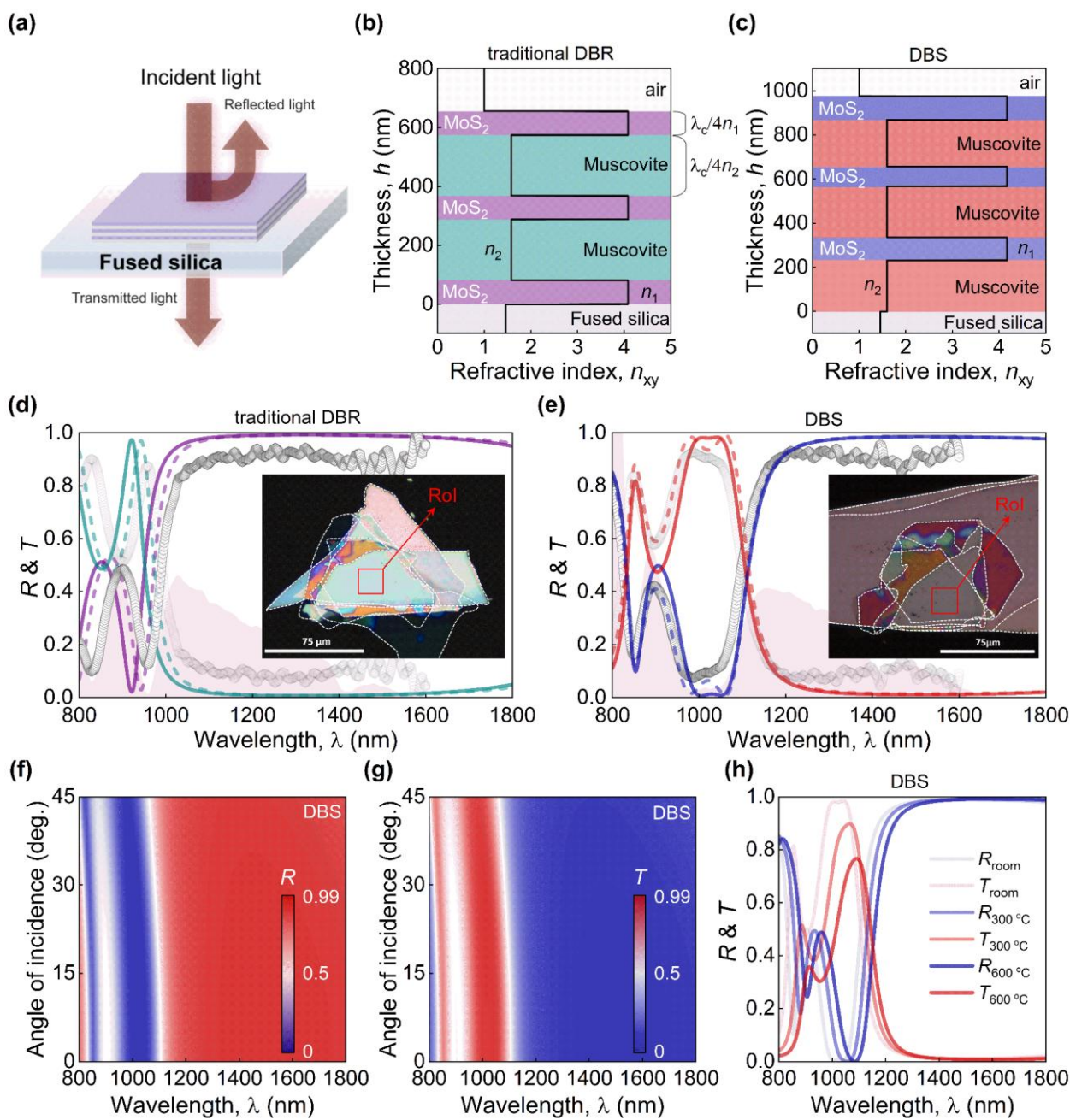

**Figure 3| Van der Waals technologies for ultrathin broadband NIR spectral region DBR and DBS nanophotonic components. a,** Schematic illustration of a few layer muscovite-MoS$_2$ heterostructure for all-vdW DBR and DBS components. **b,** and **c,** The explicit design of a five (six (c)) layer heterostructure used for an engineering of traditional (enhanced (c)) DBR (DBS (c)) photonic components with central wavelength of $\lambda_c$ = 1310 nm. The refractive indices are presented for the same central wavelength in both (b) and (c) designs. **d,** The performance characteristics of NIR spectral region broadband DBR based on a design presented in (b). **e,** The performance characteristics of NIR spectral region broadband DBS based on a design presented in (c). The solid lines in (d) and (e) show the simulated curves along with the symbols and dashed curves corresponding to the experimental data and

the fitting. The insets show 50 X optical micrographs of representative all-vdW stacks. The RoIs emphasize signal acquisition areas. The shaded areas at the background of (d) and (e) indicate stack's explicit signals acquired in the experiment. **f,** and **g,** Simulated performance alterations of all-vdW DBS with variation of an angle of incident of light. **h,** Simulated performance of all-vdW DBS at elevated temperatures up to 300 and 600 °C.

Furthermore, in the optical design (see Figure 3(c)) of short-pass all-vdW DBS operating at a same targeted wavelength, we intentionally deviate from quarter-wave condition to achieve strong transmission-band optimizing the layer thicknesses by minimizing the figures of merit. At each iteration, the optical response was computed using a generalized transfer-matrix method (see Methods) yielding an optimized six-layer all-vdW DBS exhibiting high-transmission band (average > 97 %) in 990–1050 nm spectral region and high-reflectance band (average > 96 %) in 1250 to 1800 nm with a cut-off wavelength of 1120 nm (see Figure 3(e)). The fabricated 975 nm thick prototype assembled (the individual thicknesses of the composing layers are listed in Table S4) using the same deterministic dry-transfer technique is shown in the inset of Figure 3(e) along with the measured and fitted spectra across 800–1600 nm spectral region presented within its main figure. Overall, both prototypes demonstrate good agreement with a performance comparable to the commercial DBRs and short-pass DBSs, though achieved with ultrathin sub-micron stacks composed of reduced number of layers (minor spectral shifts observed are attributed to small deviations in composing layer thicknesses). The angle-dependent reflectance and transmittance simulations display that the optical response of our all-vdW DBS remains largely unchanged up to incidence angles of 45° for an unpolarized light (see Figure 3(f-g)), which is confirmed by micro-reflectance measurements of DBS prototype at AoI of 45° in 700–1000 nm spectral region for both *s*- and *p*-polarizations (see Figure S7). Next, we further compute elevated temperature (up to 600 °C) performance of our all-vdW DBS (see Figure 3(h)) displaying a reduction and narrowing of the high-transmission band (only) highlighting the importance of red-shifts of the absorption onsets emerged in vdW $MoS_2$ with the temperature increment.

**Discussion**

An accurate optical design allowing robust operation of devised ultrathin DBR and DBS components at an elevated temperatures requires a consideration of the small amendments in material characteristics for the both composing layers, which we extract by performing temperature dependent Vis micro-reflectance studies (see Figure 4(a) and Methods) of individual vdW muscovite and $MoS_2$ samples at temperatures from room up to 600 °C. Figures 4(b) and 4(c) present the corresponding *DRC* maps for 1476 nm thick muscovite and 67 nm $MoS_2$ samples placed on fused silica substrates and acquired with unpolarized light, where the former was treated as a uniaxial crystal. Here, temperature-dependent optical response alterations emerge as Fabry–Pérot interference extrema red- shifts in thick samples of muscovite, while bearing no apparent effect in thinner ones (see Figure S8). In contrast, in $MoS_2$, the effect of temperature manifests as a gradual fading of critical points associated with *A*, *B* and *C* excitonic absorptions affecting its NIR spectral tail. These alterations of optical responses corroborate for a material based dominant-contribution between the mechanisms of various physical origins: an out-of-plane thermal expansion in muscovite and optical dispersion modification in $MoS_2$. The out-of-plane thermal expansion coefficients of muscovite family crystals typically[36] range in $10–30 \times 10^{-6}$ °C$^{-1}$, while $MoS_2$ exhibits values smaller by an order of magnitude[37]. Based on these observations, we fitted the temperature-resolved thickness variation of muscovite using the measured optical response (see Figure 4(d)), which yields to $15 \times 10^{-6}$ °C$^{-1}$ expansion coefficient. On the other hand, neglecting the latter in $MoS_2$ and using Varshni's formalism for an elucidation of temperature dependence of excitonic contributions (see Methods and Figure S9),[38–

[40] we obtain real and imaginary parts of its dielectric functions presented in Figure 4(e) with altered temperature dependent NIR spectral tails consistent with Kramers–Kronig relations and used for the simulations presented in Figure 3(h).

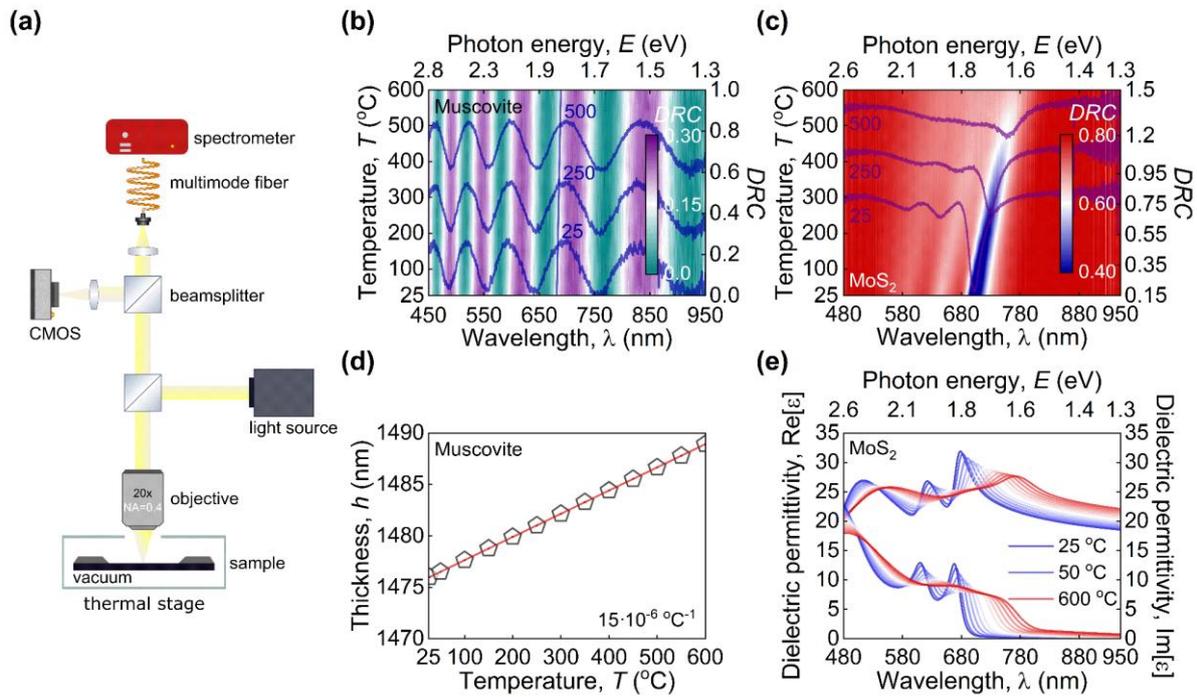

**Figure 4 | The impact of elevated temperatures on dielectric response of vdW muscovite: the dominance of thermal expansion over the optical dispersion modifications. a,** Schematic illustration of experimental setup employed for temperature-dependent spectroscopy. **b,** and **c,** DRC contourplot as function of wavelength and temperature for a representative micro-mechanically cleaved 1476 nm (67 nm) thick muscovite ($MoS_2$) sample placed onto fused silica substrate and measured in 450–950 nm (480–950 nm) spectral region at 25–600 °C temperature range. Blue (violet) curves indicate selected DRC spectra at several temperatures visualizing the evolution of critical points (Fabry–Pérot in the case of muscovite and excitonic in the case of $MoS_2$) with temperature. An offset of 0.3 (0.5) is introduced in (b) ((c)) for clarity. **d,** Out-of-plane thermal expansion of muscovite extracted from (b) considering negligible modifications of optical dispersion. **e,** The temperature evaluation of real and imaginary parts of in-plane components of dielectric permittivity function of $MoS_2$ extracted from (c).

**Methods**

**Sample preparation.** Bulk muscovite (($K^+$,$Na^+$)$Al_2$($AlSi_3O_{10}$)($OH$)$_2$) and $MoS_2$ crystals were purchased from Sigma-Aldrich and 2D Semiconductors and micro-mechanically cleaved onto standard Si, Si/$SiO_2$ and fused silica substrates at room temperature using commercial scotch tapes from Nitto Denko Corp.. Prior to the micro-mechanical cleavage, the substrates were sequentially cleaned in acetone, isopropanol and deionized water, which was followed by an air plasma treatment to remove residual surface contaminants and improve the adhesion. The multilayer vdW muscovite/$MoS_2$ heterostructures were assembled using a commercial PDMS stamp based deterministic dry transfer technique[41].

**Polarization-resolved Raman spectroscopy.** Raman spectra from vdW muscovite samples were acquired using a custom confocal microscope equipped with a 100X objective lens (*N.A.* = 0.70), Princeton Instruments SpectraPro HRS-750 triple grating spectrograph and a 1800 lines/mm (450-850 nm) diffraction grating. The measurements were carried out using a 532nm excitation laser, RazorEdge LP03-532RE-25 long pass edge filter and a Princeton Instruments PyLoN type detector operating at -75 °C temperature. The initial position of vdW muscovite samples were pre-aligned with one of the crystallographic axes. The polarization angle was varied continuously from 0° to 360° with a step of 5°.

The acquisition time was set to 1s per point with 30 accumulations. The laser power on vdW muscovite samples were maintained at 0.6 mW using a variable neutral density filter. Spectral processing involved baseline subtraction. The extracted peak intensity values were then used to construct the angular dependent polar diagrams for each of the modes. For the fitting of the latter, a slight misalignment between the crystallographic axes and the laser polarization was accounted for by an introduction of a small rotation offset of $\vartheta_0$~2.5°.

**Spectroscopic micro-ellipsometry.** Spectroscopic micro-ellipsometry measurements were carried out at room temperature over 250–1000 nm spectral region using a Park Systems Accurion EP4 rotating-compensator imaging ellipsometer. All optical modelling and fitting procedures were performed using the EP4Model software package. Multiple do muscovite samples of varying thicknesses on different substrates (Si, Si/SiO$_2$ and fused silica) were measured and analysed to ensure the reliability and accuracy of the extracted optical parameters. For each sample, the in-plane crystallographic axes were aligned with respect to the plane of incidence to enable initial characterization using standard micro-ellipsometry. In this configuration, the ellipsometric parameters $\Psi$ and $\Delta$ were collected at angles of incidence (AoIs) of 55°, 60°, and 65°. Prior to these measurements, single-point MM micro-ellipsometry was performed to verify proper alignment, confirming negligible off-diagonal element contribution. Notably, in the case of thicknesses below 250 nm, the low in-plane anisotropy leads to nearly indistinguishable $\Psi$ and $\Delta$ responses along orthogonal in-plane directions, allowing conventional ellipsometry measurements to be performed without strict alignment of crystallographic axes. To accurately extract the dielectric tensor components, extensive MM micro-ellipsometry measurements were performed on arbitrarily oriented vdW muscovite samples at 60° AoI over the 400–900 nm spectral region. The dielectric permittivity tensor components were obtained by fitting the data with the initial parameter set refined through the analysis of multiple samples and by minimizing the mean-squared error (MSE). Additional single-point MM measurements at a wavelength of 532 nm were carried out at 60° AoI while rotating the sample in 0° to 180° range with a step of 15°. This measurement sequence reduces parameter correlations and improves sensitivity to the out-of-plane dielectric response. The employed anisotropic optical model consisted of two in-plane and one out-of-plane dielectric tensor components. The dispersion of these components was described using the Cauchy relation. To account for an absorption at shorter wavelengths, the in-plane components were further parameterized using Lorentz oscillators. The optical constants of vdW muscovite were eventually obtained by minimizing the MSE of optimized anisotropic models accounting for parameter correlations and depolarization effects (see Figures S2–S5 and Table S1). The thicknesses of individual layers in all-vdW muscovite/MoS$_2$ heterostructures were extracted by spectroscopic micro-ellipsometry studies, verified sequentially after each assembly step for both DBR and DBS photonic components.

**Temperature dependent or polarization resolved Vis micro-reflectance spectroscopy.** Micro-reflectance spectra of vdW muscovite and MoS$_2$ samples were measured under normal incidence using a Leica DM6M upright optical microscope equipped with a halogen light source. Polarization resolved measurements for validation of dielectric permittivity tensor components of vdW muscovite were performed at ambient conditions, while temperature-dependent measurements for the averaged dielectric responses of vdW muscovite and MoS$_2$ at 25–600 °C range were carried out using a CRYO600-190 optical thermal stage equipped with a heating system and a vacuum pump. Spectral acquisition was performed using a Thorlabs CCS200/M spectrometer (200–1000 nm), coupled *via* a 105 μm core-diameter optical fibre. All measurements were performed on a fused silica substrate for both muscovite and MoS$_2$ samples due to its low temperature coefficient of the refractive index[42]. The differential reflectance contrast (*DRC*) was calculated as $DRC = (J_s - J_{sub})/(J_s + J_{sub})$, where $J_s$ and $J_{sub}$ are the reflected intensities from the sample and the substrate. The complex dielectric permittivity function of $MoS_2$ was obtained from $DRC$ spectra by modelling its imaginary part using Tauc–Lorentz oscillator formalism, with initial parameters derived from room-temperature spectroscopic micro-ellipsometry. Next, the real part was then calculated using a fast Kramers-Kronig relations transformation including a high-energy tail correction. Temperature-dependent $DRC$ spectra were fitted simultaneously by describing the temperature dependence of the oscillator energy $E_0$ using Varshni's relation, while the broadening parameter was modelled using Bose–Einstein's formalism associated with phonon population. Here, the optical bandgap, $E_g$, was defined relative to the oscillator energy, whereas the oscillator amplitudes were treated as free fitting parameters. The model parameters were determined by minimizing the residual between the calculated and measured spectra using a trust-region reflective nonlinear least-squares algorithm. The resulting dielectric permittivity function of vdW $MoS_2$ was extrapolated from 950 nm up to 1800 nm using the fitted model parameters (parameters at 600 °C are given in Table S5; extrapolated optical constants are shown in Figure S10).

The measured micro-reflectance spectra were modelled using a generalized 4 × 4 transfer-matrix method (TMM)[43], in which electromagnetic wave propagation and boundary conditions were explicitly treated at each interface of the optical stack. To account for the finite numerical aperture (NA) of the objective lens (NA = 0.4 for 20 X and NA = 0.5 for 50 X objectives), the illumination was described as a distribution of AoIs rather than a single angle. Consequently, the measured reflectance represented a superposition of contributions from different AoIs. This effect was incorporated by evaluating the reflectance as a weighted sum over discretized AoIs:[44] $R(\lambda) = \text{Sum}[R(\lambda, \alpha_n) W_{tot}(\alpha_n)]$, where $\alpha_n$ are the discretized AoIs, $N$ is the number of angular sampling points, and $W_{tot}$ describes the angular intensity distribution. A value of $N = 20$ was found to be sufficient for the moderate NA conditions considered. The total weighting function includes a geometric factor determined by the NA and an objective-dependent term accounting for intensity inhomogeneities within the illumination cone, modelled as: $W_{obj}(\alpha_n) = \exp[-1/2\ (\xi(2n-1)/N)^2]$, where $\xi$ is an objective-specific parameter[44]. The value of $\xi$ was determined by minimizing the difference between simulated and measured reflectance spectra of a $SiO_2/Si$ reference sample with known thickness, ensuring accurate reproduction of the experimental response (see Figure S11). Notably, moderate NAs introduce only minor corrections in the case of thin layers, but become increasingly important for thicker samples, as demonstrated in Figure S12.

**NIR micro-reflectance spectroscopy.** NIR spectral region micro-reflectance spectroscopy measurements of all-vdW DBR and DBS components were performed under a normal incidence using the same optical setup described above, equipped with a halogen light source covering 800–1600 nm spectral region and coupled to an Ando AQ6315A optical spectrum analyser *via* a 105 µm core-diameter optical fibre. Measurements were carried out using a 50 X objective, enabling spectral acquisition from regions with lateral dimensions of approximately 4 µm. The reflectance was calculated as $R = R_{ref} \times (J_s - J_{dark})/(J_{ref} - J_{dark})$, where $J_s$ is the signal measured from the sample, $J_{ref}$ is the signal measured from the reference mirror, and $J_{dark}$ is the detector dark signal acquired with the light source blocked accounting for the background and instrumental noise. The reference was a protected aluminium mirror (Thorlabs PFSQ10-03-G01). The reference reflectance data were provided by Thorlabs for an AoI of 6°, which closely approximates the normal-incidence conditions for metallic mirrors. The transmittance was then estimated as $T = 1 - R$ assuming negligible absorption in $MoS_2$ over the investigated spectral region. Micro-reflectance measurements were additionally performed using the Accurion EP4 imaging spectroscopic ellipsometer to probe the performance of all vdW DBS components at an AoI of 45° over 700–1000 nm spectral region, employing a customized Python script for data acquisition. The measured micro-reflectance spectra were then modelled using a generalized 4 × 4 TMM, in which electromagnetic wave propagation and boundary conditions were explicitly treated at each interface of the optical stack.

**All-vdW DBR and DBS calculations.** The optical responses of our all-vdW muscovite/$MoS_2$ stacks for DBR and DBS configurations were calculated for an unpolarized light at a normal incidence (AoI = 0°) using a generalized 4 × 4 transfer-matrix formalism[43]. In the case of a DBR, the alternating high- and low-refractive index layers ($MoS_2$ and muscovite) were designed using the quarter-wave optical thickness condition, $h_H = \lambda_c/4n_H$, $h_L = \lambda_c/4n_L$, where $n_H$ and $n_L$ are the refractive indices of our high-index and low-index materials. This condition ensures constructive interference of reflected waves, resulting in a high reflectance stop band around $\lambda_c$ (1310 nm for the present design). In the case of DBS, the design was based on controlled deviation from the quarter-wave condition. Here, the initial layer thicknesses were defined using the quarter-wave optical thickness condition and subsequently tuned to achieve the targeted spectral performance. The optimization targeted high transmission and low reflection around 1030 nm, and high reflection and low transmission around 1400 nm. Specifically, the figure of merit penalized deviations from $T \geq 0.98$, $R \leq 0.02$ around 1030 nm, and $R \geq 0.98$, $T \leq 0.02$ around 1400 nm, evaluated over finite spectral windows (± 30 nm). A soft constraint on total stack thickness (< 1 µm) was also included. It is important to note, that the refractive index of vdW muscovite is quite close to that of the fused silica substrate, therefore, the first ~235 nm thick muscovite layer (adjacent to the substrate) can be omitted without significant impact on the performance characteristics. The optimization was performed using derivative-free methods (bounded Powell search[45] followed by Nelder–Mead refinement[46]), and final reflectance/transmittance spectra were calculated across the Vis–NIR spectral region. For the design of these heterostructures, the dielectric permittivity of $MoS_2$ used in the computations was independently determined by spectroscopic micro-ellipsometry in the 250-1000 nm spectral region (see Figure S10) and extrapolated up to 1800 nm together with vdW muscovite. VdW muscovite was modelled as a uniaxial crystalline layer using a single in-plane dielectric tensor component in the case of both DBR and DBS components as calculations using orthogonal in-plane components displayed negligible differences (see Figure S13).